# Spin Waves and Spin Currents in Magnon-Phonon Composite Resonator Induced by Acoustic Waves of Various Polarizations

S.G. Alekseev[a,*], N.I. Polzikova[a], V.A. Luzanov[b], S.A. Nikitov[a]

[a] *Kotelnikov Institute of Radioengineering and Electronics RAS*
*Mokhovaya Str. 11, Build. 7, Moscow 125009 Russian Federation*
[b] *Fryazino branch Kotelnikov Institute of Radioengineering and Electronics RAS*
*Vvedenskogo Squar. 1, Fryazino, Moscow region 141190 Russian Federation*
*E-mail: alekseev@cplire.ru

***Abstract.*** In this work, we present the results of a systematic experimental study of linear and parametric spin wave resonant excitation accompanied by spin currents (spin pumping) in a multifrequency composite bulk acoustic wave resonator with a ZnO-YIG-GGG-YIG/Pt structure. The features of magnetic dynamics excitation in YIG films due to magnetoelastic coupling with acoustic thickness modes of various polarizations are studied. Acoustic spin waves and spin pumping are detected by simultaneous frequency-field mapping of the inverse spin Hall effect voltage and the resonant frequencies of thickness extensional modes. In the parametric range of frequencies and fields, acoustic spin pumping induced by both shear and longitudinal polarization modes was observed. Linear acoustic spin waves are excited only by shear thickness extensional modes because longitudinal acoustic waves do not couple with the magnetic subsystem in linear regime.

## Introduction

Magnon-phonon interactions determine the fundamental properties of magnetic materials and structures, such as relaxation processes, and are also of practical interest, for example, for low-energy consumption for spin waves (SW) and spin currents excitation [1-5]. In composite heterostructures containing piezoelectric and ferro(ferri)magnetic layers, the excitation of so-called acoustic SW (ASW) and acoustic spin pumping (ASP) occurs due to a combination of magnetoelasticity and piezoelectric effect in various layers, not necessarily in direct contact. To generate ASW, both surface acoustic waves (AW) excited by interdigitated transducers [1, 4-9] and volume AW, in particular, microwave modes in composite High overtone Bulk Acoustic wave Resonator (HBAR) [3, 10, 11] are used. Currently, HBARs, along with surface AW resonators, have proven to be in great demand as sources of coherent phonons for fundamental and applied research [12-15].

In our previous works, we studied phenomena associated with the interaction of coherent AW and SW in hybrid magnon-phonon HBARs with a layered structure: piezoelectric (ZnO) – ferrimagnetic (yttrium iron garnet - YIG) – dielectric substrate (gallium gadolinium garnet - GGG) – ferrimagnetic (YIG) – Pt (Fig.1) [16-21]. In the structure involved, shear bulk acoustic thickness modes of high harmonics were excited using piezoelectric transducers in the gigahertz frequency range. A self-consistent theory was developed to describe magnetoelastic phenomena in such structures [16-18]. The acoustic excitation of both linear [16-19] and parametric ASWs [20, 21] and their electrical detection via the effect of spin pumping and the inverse spin Hall effect (ISHE) were theoretically proved and experimentally demonstrated.

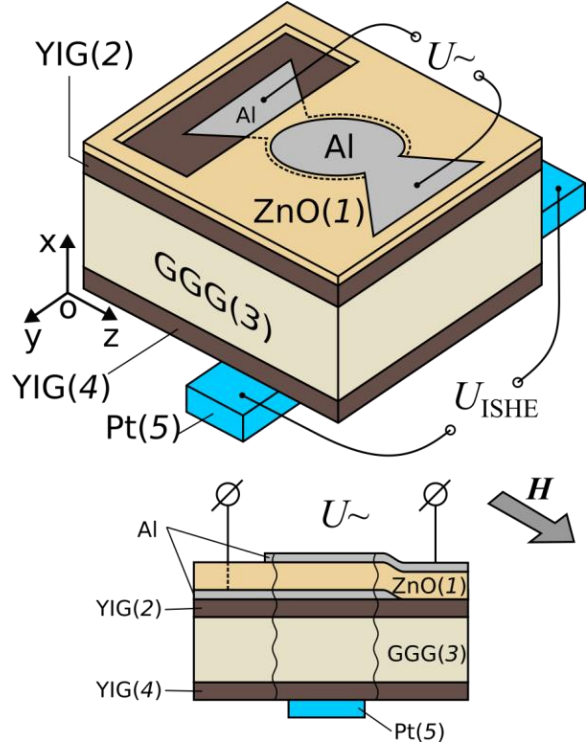

Fig. 1. Schematic of hybrid HBAR

In this work, we present the experimental study of linear and parametric ASWs excitation and the features of the spin pumping they create in the hybrid magnon-phonon HBAR due to thickness acoustic modes of various polarizations: transverse (shear) and longitudinal. As in our previous works, we use the method of acoustic resonator spectroscopy [22] in combination with the method of electrical detection of ISHE voltage. In particular, the frequency-field ($f,H$) dependences of the microwave signal complex reflection coefficient $S_{11}(f,H)$ from the transducer electrodes and the constant voltage $U_{\text{ISHE}}(f,H)$ on a platinum strip are studied.

## 1. Methods

Experimental hybrid HBARs (see Fig. 1) are fabricated based on ready-made structures consisting of a 500 μm thick (111)-oriented GGG substrate (*3*) and a 30 μm thick epitaxial YIG films (*2*), (*4*). The YIG films were doped with La and Ga. Piezoelectric transducer composed of ZnO film (*1*) sandwiched between two thin-film aluminum electrodes was deposited on one side of the structure by rf magnetron sputtering. The top and the bottom electrodes were patterned by photolithography and had an overlap with the aperture $a = 170$ μm. The Pt film (*5*) was deposited onto the free film (*4*) and formed as a stripe. The HBAR technology and design are described in more detail in [18-20, 22].

Electrical excitation and detection of bulk AWs of different polarizations occur due to the direct and reverse piezoelectric effect in a ZnO film with an inclined $\vec{c}$- axis [22, 23]. Depending on the magnitude of the applied magnetic field, ADSW excitation in YIG films due to magnetoelastic interaction takes place either at HBAR frequencies $f_n$ (linear regime) or at half frequencies $f_n/2$ (parametric regime) when the threshold power is exceeded. The spin current from YIG into Pt $\vec{j}_s$, [24] created by ADSW, is converted into conductivity current by ISHE [25]. This results in a constant voltage detected at the ends of a platinum thin film strip

$$U_{\text{ISHE}} = -a'(\vec{E}_{\text{ISHE}} \cdot \vec{y}), \qquad \vec{E}_{\text{ISHE}} \propto -(\vec{j}_s \times \vec{z}). \qquad (1)$$

Here $\vec{E}_{\text{ISHE}}$ is an electrostatic field, $a' \approx a$ is the length of the region in the *y* direction in which ADSW excitation takes place.

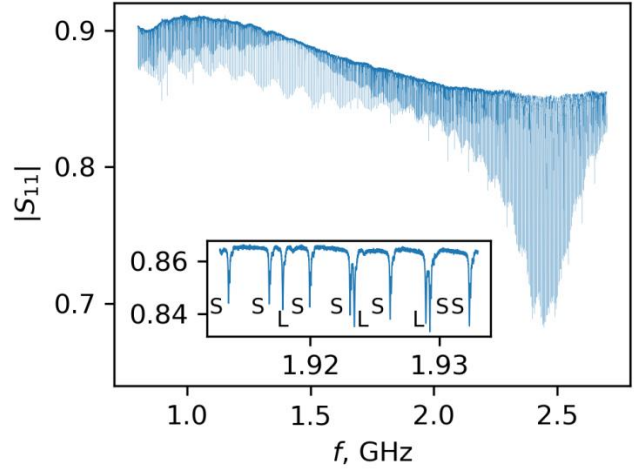

Fig. 2. Frequency dependence of the microwave reflection coefficient modulus $|S_{11}(f)|$ in the absence of a magnetic field. The inset shows an enlarged fragment. The dips in the frequency response correspond to the resonant frequencies of the thickness modes of shear AW (S) and longitudinal AW (L); the intermodal distance of longitudinal modes is approximately twice as large as that of shear modes.

## 2. Results and discussion

Figure 2 shows the frequency dependence of reflection coefficient modulus $|S_{11}(f)|$ of microwave signal from the piezoelectric transducer electrodes in the absence of a magnetic field. All the experiments were conducted at fixed power level 9 mW. To study the excitation features of ASW and ASP from acoustic modes of different polarizations, the frequency range corresponding to the inset in Fig. 2 was selected. In this range the transducer excites both longitudinal (L) and shear (S) modes with the same efficiency. We denote the frequencies of these modes as $f_l^L$ and $f_s^S$, where $l$ and $s$ are the overtone numbers of the thickness modes of the corresponding polarizations. Further studies are carried out in a tangential magnetic field $H$ in the range (0 – 450 Oe). As will be shown below, this field range contains both linear and parametric regimes of the ASW excitation [20].

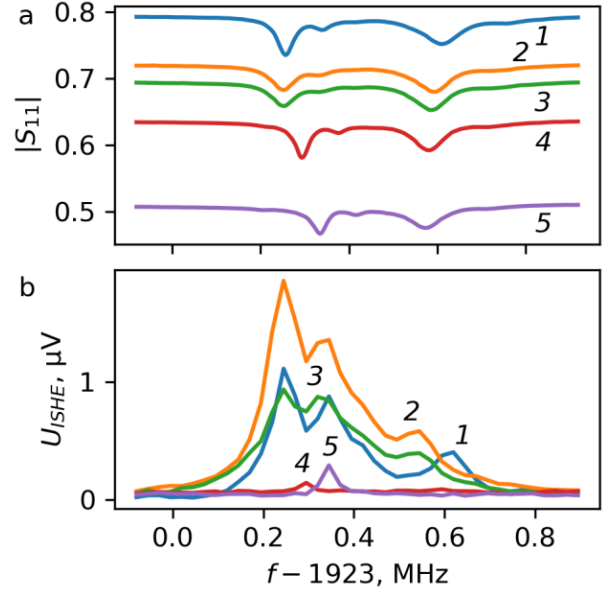

Fig. 3. Frequency dependencies of $|S_{11}(f)|$ (a) and voltage $U_{ISHE}(f)$ on Pt (b) at several magnetic fields. Curves: *1* – 60 Oe, *2* – 181 Oe, *3* – 238 Oe, *4* – 327 Oe, *5* – 352 Oe.

Figure 3a shows the frequency dependences of the reflection coefficient (Fig. 3a) and the ISHE voltage $U_{ISHE}$ (Fig. 3b) measured simultaneously at several magnetic fields. The measurements were carried out in a narrow frequency range, including closely located one longitudinal and one transverse AW modes (see inset in Fig. 2).

As one can see from Fig. 3a, the resonant frequency for the longitudinal mode $f_l^L(H)$ (1923.6 MHz at $H$=0) changes slightly with the field increase. The resonant frequency of the transverse mode $f_s^S(H)$ (1923.3 MHz at $H$=0) remains practically unchanged in weak fields and experiences a shift in the fields $H > 200$

Oe. The shift increases as the field approaches the ferromagnetic resonance (FMR) region. Assuming $f_s^S \approx f_{FMR}$, where the FMR frequency is related to the magnetic field by the Kittel formula

$$f_{FMR} = \gamma[H(H+4\pi M_0)]^{1/2}, \qquad (2)$$

we find that $H_{FMR} \approx 384$ Oe. Here, $\gamma = 2.8$ MHz/Oe, $M_0$ - effective saturation magnetization. For doped YIG we use the value $4\pi M_0 = 845$ Oe, established for an identical structure in [20].

The change in the positions of voltage maxima $U_{ISHE}(f)$ upon excitation of the transverse mode demonstrates similar behavior in the fields $H > 200$ Oe, but significantly more diverse behavior at lower fields. Figure 3b clearly shows that the $U_{ISHE}$ maximum splits into two. At the same time on the characteristics $|S_{11}(f)|$ in Fig. 3a there is a mild feature: a minimum located at 80 kHz higher from the main one and corresponding to the splitted $U_{ISHE}$ maxima mentioned above.

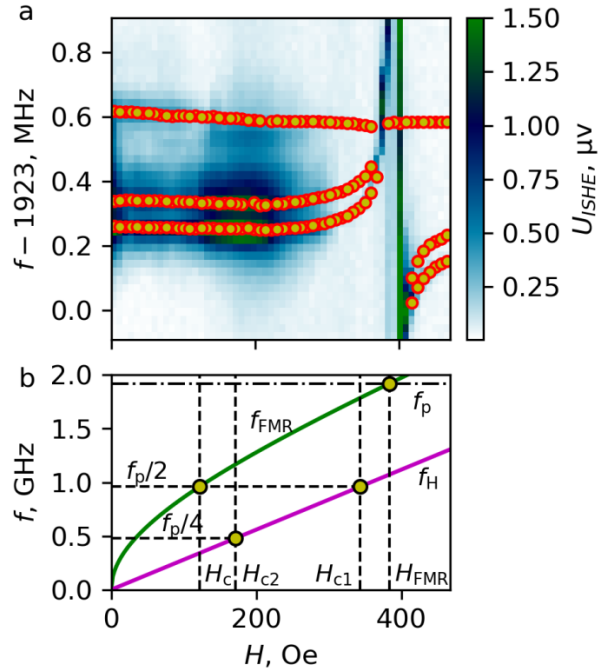

Fig. 4. Frequency-field dependence of the voltage $U_{ISHE}(f,H)$ (a). The dots show the minimums of the $S_{11}$ reflectance. The field dependences of the SW spectra frequency limits (b).

For detailed comparison of the behavior $|S_{11}(f, H)|$ and $U_{ISHE}(f,H)$ let us present them on the same graph. For this, the 3D color map $U_{ISHE}(f,H)$ (Fig. 4.a) is best suited, on which the minima $|S_{11}(f, H)|$, (dots) are superimposed. Also let us consider the magnetic field dependencies mentioned above in accordance with the calculated dependences of the SW spectra frequency limits $f = f_H(H) = \gamma H$ and $f = f_{FMR}(H)$ shown in Fig. 4b. The horizontal lines mark the frequency $f_p$=1.1923 GHz $\approx f_{s,l}^{S,L}$, and its sub-harmonics $f_p/2$ and $f_p/4$. The critical fields marked in Fig. 4b are found from the relations

$$H_{FMR} = [(2f_p/\gamma)^2+(4\pi M_0)^2]^{1/2}/2 - 2\pi M_0 = 384 \text{ Oe}, \quad H_{c1} = f_p/(2\gamma) = 343 \text{ Oe},$$

(3)

$$H_c = [(f_p/\gamma)^2+(4\pi M_0)^2]^{1/2}/2 - 2\pi M_0 = 121 \text{ Oe}, \quad H_{c2} = f_p/(4\gamma) = 171.5 \text{ Oe}.$$

The linear excitation of ADSW results in the signal of $U_{ISHE}(f, H)$ in the vicinity of the $H_{FMR}$ field. Additional non-resonant (i.e. frequency independent) contributions to the $U_{ISHE}$ signal (Fig. 4a) and to the decrease in the overall level of $|S_{11}(f)|$ at $H \approx H_{FMR}$ are associated with inductive excitation of magnetic dynamics directly by the transducer electrodes. Such mixed inductive and acoustic excitations, as well as the possibility of completely acoustic excitation of SW, were discussed in detail in [19].

The excitation of any parametric SW is possible if $H < H_{c1} = f_p/(2\gamma)$. Therefore, in the field range $H_{c1} < H < H_{FMR} \approx H_{MER}$, only linear excitation of ADSW is possible due to the magnon – transverse phonon coupling. Here the field $H_{MER}$ is the field of magnetoelastic resonance, at which synchronism between SW and AW occurs, $H_{MER}(f)=H_{FMR}+H_{ex}$, where $H_{ex} \sim 3 - 5$ Oe is the field of inhomogeneous exchange [18]. It can be seen that in the linear field region there is a direct match between the voltage maxima position and the main resonant frequency of the shear mode.

It can be noted also that both transverse AW modes induce voltage $U_{ISHE}$ in the parametric region ($H < H_{c1}$), and the signal maximum is located in the region $H_c < H < H_{c2}$. The field $H_c$ corresponds to the creation process of two parametric magnons with frequency $f_p/2$ and zero momentum, and $H_{c2}$ corresponds to the upper limit on $H$ for the possible decays of parametric magnons with the frequency $f_p/2$ into two secondary parametric ones at a frequency $f_p/4$. A more detailed discussion of the critical fields given in (3) see [19].

Let's consider the case of longitudinal mode. As can be seen from Fig. 4 a, the longitudinal AW mode does not affect $U_{ISHE}$ in the linear regime, and in the parametric one its influence is limited by the fields $H < H_c = 121 Oe$. Note that a small $U_{ISHE}$ signal is also detected in fields $130 < H < 280$ Oe, but at excitation frequencies that do not correspond to either the L or S HBAR modes. This is clearly visible, for example, from a comparison of curves *2* in Fig. 3a and Fig. 3b. The reason for this response is not yet clear. Note that parametric spin pumping induced by both S and L modes at frequencies of about 2.4 GHz was also observed in [19].

Note that the presence of additional resonant frequencies in the HBAR spectrum is due to locality of the elastic oscillations excitation. The excitation region in the structure plane is determined by the transducer aperture with diameter *a*. Strictly speaking, these oscillations will propagate not only under the transducer, as shown in Fig. 1, but also outside it, carrying energy away from the excitation region in the form of plate Lamb modes [26, 27]. The highest resonator quality factor is achieved when the so-called trapped-energy regime is realized. Namely, at a certain ratio of frequencies and geometric dimensions, there are no conditions outside the transducer region for propagating modes. In this case, the elastic energy remains localized in the transducer region with an energy distribution decreasing exponentially with distance from the electrode edge. In this region fundamentally trapped-energy high overtone thickness modes are quasi-uniform in plane. In addition to fundamental modes, one or more lateral standing modes may be exited. These modes, which are also trapped-energy, are located higher in frequency from the fundamental ones and are usually called spurious resonance [26].

In our case, at least a small spurious S-mode is observed near the main one at a frequency of 1923.3 MHz. It can be noted that the depths of the $|S_{11}|$ dips for the main and the spurious modes differ several times (Fig. 3a). However, the heights of the corresponding resonant peaks on the $U_{ISHE}$ are comparable (Fig. 3 b, Fig. 4). Such inconsistency can be explained as follows. The ISHE voltage according to (1) depends on the $E_{ISHE}$ field magnitude, which is obviously greater for the fundamental mode. As for the length of the spin pumping region $a'$, it turns out to be larger for the spurious mode compared to the main one, for which $a' = a$, since spurious mode is less localized near the electrode boundaries. In this way, partial compensation occurs for the acoustic energy attributable to the non-fundamental mode.

## Conclusion

The electroacoustic excitation of magnetic dynamics in YIG films in a magnon-phonon bulk acoustic resonator has been studied. The regimes of linear and parametric spin waves and spin currents excitation due to thickness extensional modes with various polarizations have been studied. In the linear regime, spin dynamics in the YIG films is excited only by transverse modes (both fundamental and spurious thickness overtones). In the parametric regime (in lower magnetic fields), the spin dynamics in YIG films is excited by acoustic modes of various polarizations (both transverse and longitudinal).

The authors declare no conflicts of interest.

## Funding

This work was carried out in the framework of the State task "Spintronics-2".

## List of References